\documentclass[aps,prb,preprint,showpacs,unsortedaddress,amsmath,amssymb]{revtex4-1} 
\usepackage[english]{babel}
\usepackage[utf8]{inputenc}
\usepackage{amsmath} 
\usepackage{braket}
\usepackage{mathtools}
\usepackage{mathrsfs}
\usepackage{graphicx}
\usepackage[colorinlistoftodos]{todonotes}
\usepackage{afterpage}

\usepackage{dcolumn}
\usepackage{bm}
\usepackage{url}
\usepackage[colorlinks=true,linkcolor=blue,citecolor=blue,urlcolor=blue]{hyperref}
\usepackage[normalem]{ulem}

\usepackage{color}

\begin{document}

\title{Thermoelectric phonon glass electron crystal via ion beam patterning of silicon}

\author{Taishan Zhu}
\altaffiliation
{PhD candidate, Department of Mechanical Science \& Engineering} %
\author{Krishnan Swaminathan-Gopalan}
\altaffiliation
{PhD candidate, Department of Mechanical Science \& Engineering} %
\author{Kelly Stephani}
\altaffiliation
{Assistant Professor, Department of Mechanical Science \& Engineering} %
\author{Elif Ertekin}
\altaffiliation
{Associate Professor, Department of Mechanical Science \& Engineering} %
\email{Corresponding author: ertekin@illinois.edu}
\affiliation
{Department of Mechanical Science \& Engineering, University of Illinois at Urbana-Champaign, IL 61820}

\begin{abstract}
Ion beam irradiation has recently emerged as a versatile approach to functional materials design. 
We show in this work that patterned defective regions generated by ion beam irradiation of silicon can create a phonon glass electron crystal (PGEC), a longstanding goal of thermoelectrics. 
By controlling the effective diameter of and spacing between the defective regions, molecular dynamics simulations suggest a reduction of the thermal conductivity by a factor of $\sim$20 is achievable. 
Boltzmann theory shows that the thermoelectric power factor remains largely intact in the damaged material. 
To facilitate the Boltzmann theory, we derive an analytical model for electron scattering with cylindrical defective regions based on partial wave analysis. 
Together we predict a figure of merit of $ZT \approx 0.5$ or more at room temperature for optimally patterned geometries of these silicon metamaterials.
These findings indicate that nanostructuring of patterned defective regions in crystalline materials is a viable approach to realize a PGEC, and ion beam irradiation could be a promising fabrication strategy. 
\end{abstract}

 

\maketitle

\newpage

\section{Introduction}

Since their discovery, thermoelectric materials have attracted extensive interest for direct conversion between heat and electrical energy {\it via} Seebeck/Peltier effects \cite{majumdar2004thermoelectricity,tritt2006thermoelectric,sootsman2009new,minnich2009bulk,heremans2013thermoelectrics}. 
As opposed to fossil fuels, thermoelectrics are pollution-free during operation, stable, and have decent manufacturing scalability \cite{tritt2006thermoelectric,heremans2013thermoelectrics}. 
Nevertheless, the thermoelectric conversion efficiency must be enhanced for large-scale future adoption  \cite{majumdar2004thermoelectricity,tritt2006thermoelectric,sootsman2009new,minnich2009bulk,heremans2013thermoelectrics}. 
The conversion efficiency is given by \cite{nolas2013thermoelectrics}   
\begin{equation}
\eta=\eta_C \frac{\sqrt[]{ZT+1}-1}{\sqrt[]{ZT+1}+T_H/T_C} \hspace{0.5em},
	\end{equation} 
which, as the figure of merit $ZT=\sigma S^2 T/\kappa$ increases, approaches the Carnot efficiency $\eta_C$ of an engine operating between heat baths with  temperatures $T_H$ and $T_C$.
Here $\sigma$ is the electrical conductivity, $S$ the Seebeck coefficient, and $\kappa$ the total thermal conductivity, which aggregates contributions from electrons and phonons.  
Since $\sigma, S$, and $\kappa$ are intrinsically related material parameters, they must be carefully coordinated in order to achieve a high $ZT$. 

To this end, early efforts focused separately on either {\it thermal} or {\it electrical} properties \cite{nolas2013thermoelectrics}. 
While $\kappa$ can be reduced by phonon engineering \cite{balandin2005nanophononics}, the power factor $S^2\sigma$ can be enhanced by doping and electron band structure engineering such as in low-dimensional materials and nanostructures \cite{hicks1993effect,dresselhaus2007new}.  
In 1990s, the separate approaches were merged culminating in the notion of the phonon glass electron crystal (PGEC) \cite{slackPGEC1995}, in which a material is perceived as glassy by phonons, but remains crystalline for electrons. 
To realize a phonon glass electron crystal, several approaches have proven promising. 
First, scattering of phonons {\it via} disorder, such as by alloying, rattler structures, and point defects, has been demonstrated. 
The alloying approach recently has achieved a high $ZT \approx 2.3$ for iodine-doped Cu$_2$Se \cite{liu2013ultrahigh}. 
Second, scattering of phonons through nanostructuring such as superlattices and nanowires can also be effective. 
A $ZT \approx 2.4$ was reported in $p$-type Bi$_2$Te$_3$/Sb$_2$Te$_3$ superlattices at room temperature \cite{venkatasubramanian2001thin,harman2002quantum}. 
Third, complex crystals are now emerging \cite{Snyder2008}, including skutterudites \cite{nolas1999skutterudites} and half-Heusler alloys \cite{zhu2015high}. 
The $\kappa$ of these compounds is often below 5 W/mK, comparable to glasses, contributing to a $ZT$ around unity \cite{Snyder2008}. 
If only material performance were relevant,  these recently reported examples would already be quite competitive. 
 
However, for thermoelectric deployment at global scales, it is imperative to account for material cost and scalability of manufacturing  \cite{leblanc2014material,tritt2006thermoelectric}. 
Most notable thermoelectric materials contain elements such as Bi, Te, Sb, Pb, and Ag, that are either expensive, toxic, or challenging for processing. 
By contrast, silicon, the most widely used material, is nowadays being reconsidered as a promising candidate \cite{leblanc2014material,hochbaum2008enhanced,bux2009nanostructured,lee2008nanoporous}. 
Due to its low cost and viable manufacturability, investigations for thermoelectric applications \cite{yee2013per} in both bulk alloy and nanostructured form  \cite{leblanc2014material,bux2009nanostructured} have regained interest. 
For instance, bulk Si$_{0.98}$Ge$_{0.02}$ has an appealing $ZT\sim 0.32$ at competitive price of 1.7 US\$/Watt \cite{bux2009nanostructured}. 


In this work, we 
propose a silicon nanocomposite composed of regularly patterned defective regions embeddeded in a crystalline host, as shown in Fig. \ref{figSchem}(a), for thermoelectric applications. 
In our recent work (Ref. [\onlinecite{defects2016}]), we showed that the effective diameter of defective regions $D$ and  the spacing between them $L$ can be controlled varying the parameters of the ion beam irradiation process, such as ion type, irradiation energy, fluence, beam diameter, and beam incidence angle. 
The physical justification for the proposed metamaterial is illustrated conceptually in Fig. \ref{figSchem}(b). 
Due to the long phonon mean free paths $\Lambda_p$ in silicon relative to the corresponding electron mean free paths $\Lambda_e$, we expect that as the nanostructure feature size grows, the electronic conductivity will increase and saturate more quickly than the thermal conductivity. 
If the inter-defective region distance $L$ falls within the length window spanned by the mean free path of electrons $\Lambda_e$ and that of phonons $\Lambda_p$, then $\kappa$ can be reduced due to phonon scattering while $\sigma$ is largely retained, thereby $ZT$ can be enhanced. 



The purpose of the present work is to verify that the proposed metamaterial formed by ion beam patterning of silicon can lead to a PGEC.
Atomic simulations are performed to determine the phonon transport and thermal conductivity, while the Boltzmann transport theory is employed to estimate the electrical properties. 
Whereas Green-Kubo calculations have been well established for obtaining thermal conductivity, the scattering model for electrons with the cylindrincal defective regions is currently not available in the literature. 
To bridge this gap, we derive an analytical scattering model based on the partial wave approach. 
Using this scattering model, we demonstrate that it is possible to achieve substantial reduction in $\kappa$ without sacrificing electrical properties, and predict that $ZT \approx 0.5$ or greater is achievable. 
This compares well to other nanostructured silicon systems reported in the literature such as silicon nanowires ($ZT \approx 1$)\cite{hochbaum2008enhanced,boukai2008silicon} and nanoporous silicon ($ZT \approx$ 0.4)  \cite{lee2008nanoporous,tang2010holey},
but practically has the advantage of ease of manufacturability.



\section{Computational methods and theoretical models}

To study the thermoelectric transport properties, we used different techniques for phonons and electrons. For the prediction of thermal conductivity, we applied Green-Kubo formalism implemented in equilibrium molecular dynamics simulations. Meanwhile, for electrical properties, we resorted to Boltzmann theory and the relaxation time approximation. 

\subsection{Equilibrium molecular dynamics for $\kappa$}
\begin{figure*}[!hbtp] 
\centering
\includegraphics[width=0.9\textwidth]{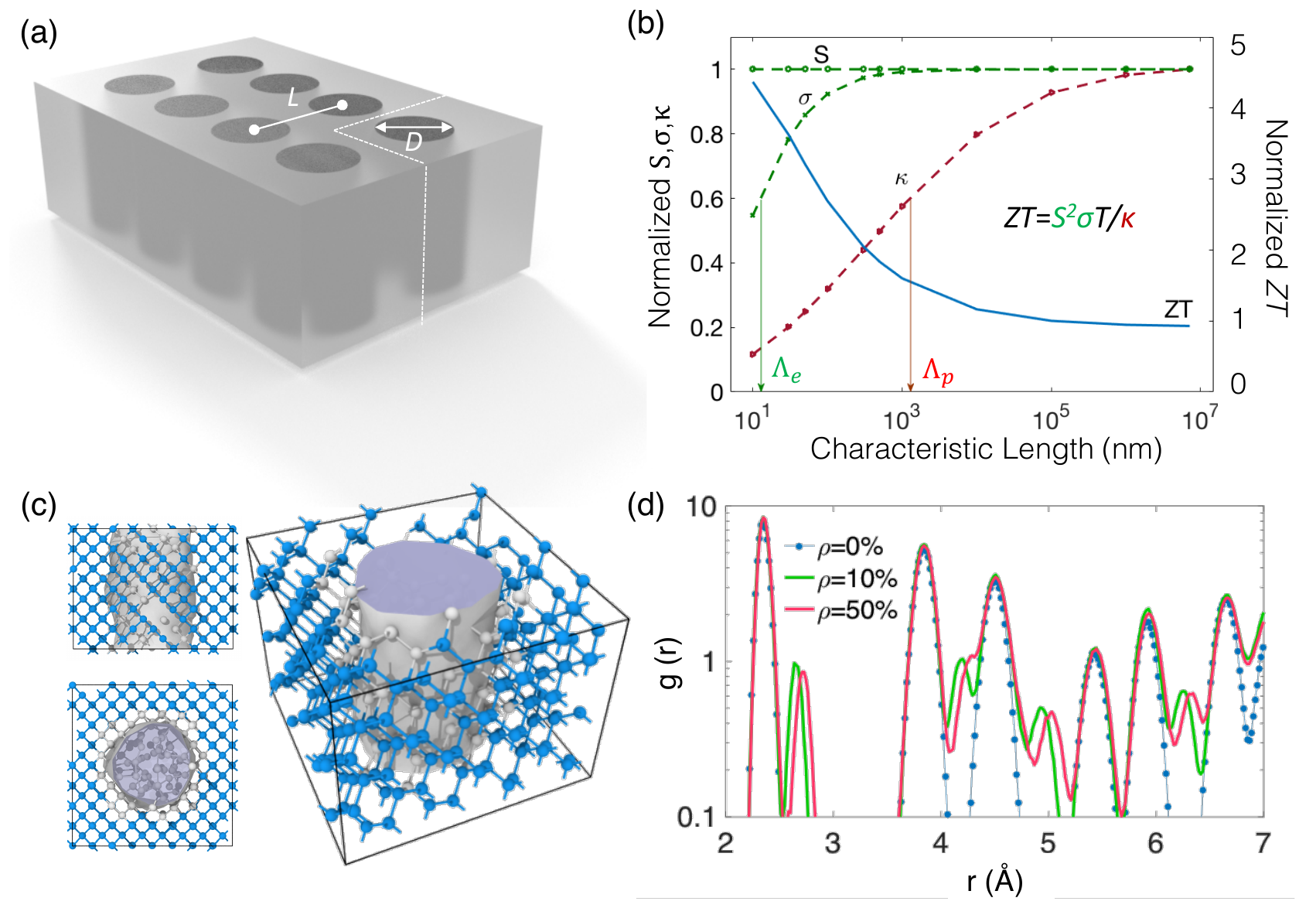}
\caption{\label{figSchem} 
(a) Schematic of silicon metamaterial for thermoelectric applications, where the dark areas denote damaged domains patterned by ion beam irradiation. 
(b) The rationale for the patterned system, in which electrical properties are expected to increase towards the bulk values faster than thermal properties with increasing feature size as a result of the different phonon and electron mean free paths. 
If the feature size is larger than electron mean free path $\Lambda_e$  and smaller than the phonon mean free path $\Lambda_p$, $ZT$ can be enhanced. 
(c) A representative super-cell for (a) in atomic view, with disordered regions generated by ion beam irradiation, obtained by molecular dynamics simulations. 
(d) The radial distribution function for the specimens shown in (a,c). New peaks and fine shifts can be observed as the degree of disorder $\rho$ increases.
}
\end{figure*}

The ion beam irradiated materials are created by direct simulation of
ion bombardments using molecular dynamics simulations, as described in detail in our previous work\cite{defects2016}. 
The impact location is randomly chosen from a two-dimensional normal distribution parameterized by beam diameter, which mimics a focused ion-beam apparatus. As annealing is expected to be most prominent in the first few picoseconds after ion impact, we allow the system to anneal for 70 ps at $T=300$K between two consecutive ion impacts. 
An ensemble of 50 independent irradiation processes are simulated to obtain satisfactory statistics. 
All molecular dynamics calculations were performed using HOOMD-blue \cite{Anderson2008}. 
The interactions between silicon atoms are described by the Tersoff potential \cite{Tersoff1988}, and ion-Si interactions by the Ziegler-Biersack-Littmark universal repulsive potential\cite{ziegler1985stopping}. 
Figure \ref{figSchem}(c) illustrates an example of a sample irradiated by a 5 keV Xe ion beam oriented normal to the surface. 
The damaged region is characterized by as a cylindrical region with diameter describing the radial extent and height describing the range of damage (see Fig. \ref{figSchem}(c)). 
The corresponding radial distribution function for varying degrees of disorder is shown in Fig. \ref{figSchem}(d), where new peaks are generated due to the presence of disorder, which drift as the degree of disorder increases.

The thermal conductivity $\kappa$ of the irradiated samples is calculated
using the Green-Kubo formulism\cite{frenkel2001MD},
which relates $\kappa$ to the fluctuation of heat flux,
\begin{equation} \label{eq_green-kubo}
\boldsymbol{\kappa} = \frac{1}{k_BVT^{2}} \int_{0}^{\infty} \left\langle\boldsymbol{J}(t) \cdot \boldsymbol{J}(0)\right\rangle d t
\end{equation}   
based on the fluctuation-dissipation theorem.
Here $k_{B}$ is the Boltzmann constant, $V$ volume, $t$ time, and $\left\langle\boldsymbol{J}(t) \cdot \boldsymbol{J}(0)\right\rangle$ the auto-correlation function of heat current $\boldsymbol{J}$ calculated from molecular dynamics simulations. 
The integral is considered converged once the statistical errors fall within 5\%.
All simulations were performed at $T = 300 \, {\rm K}$ with a time step of 0.5 fs. 
The system was equilibrated to the desired temperature for 20 ps with a Berendsen thermostat, and then sampled in the microcanonical ensemble (NVE) for an additional 20 ps. 
The heat current was then recorded for a simulation time of 6 ns. 
For each value of $\kappa$ reported below, 10 independent micro-states are simulated, and $\kappa$ is averaged over in-plane directions $\kappa = (\kappa_{x} + \kappa_{y})/2$. 
The calculated $\kappa$ of pristine silicon at room temperature is approximately 270 W/mK from this method, almost twice that of the experimentally observed value of 150 W/mK \cite{Fulkerson1968}. However, this numerical value is consistent with other molecular simulations using the same potential\cite{lee2007lattice}.

\subsection{Boltzmann theory for $\sigma, S$}

For the electronic properties $\sigma$ and $S$ we have applied Boltzmann theory. We use the relaxation time approximation and the parabolic bands approximation for the electronic dispersion. 
These approximations are sufficiently accurate for non-degenerately doped silicon, since at typical thermoelectric operating temperatures ($T = 300 \, {\rm K}$ to $T = 700 \, {\rm K}$) the filling of the conduction bands is relatively small   \cite{lundstrom2009fundamentals,ma2012thermoelectric}. 
Within this framework the kinetic definitions of $\sigma$ and $S$ are given by \cite{nolas2013thermoelectrics} 
\begin{eqnarray}
\label{sigmaBoltz}
\sigma &=& -q^2 \int v(\epsilon)^2 \tau(\epsilon) \frac{\partial f}{\partial \epsilon} g(\epsilon) \, d\epsilon   \hspace{0.5em}, \\
\label{SBoltz}
S &=& \frac{1}{qT} \frac{\int  v(\epsilon)^2 \tau(\epsilon) \frac{\partial f}{\partial \epsilon} [\epsilon -\mu] g(\epsilon) \, d\epsilon }{\int v(\epsilon)^2 \tau(\epsilon) \frac{\partial f}{\partial \epsilon} g(\epsilon) \, d\epsilon }  \hspace{0.5em},
\end{eqnarray}
where $q$ is the elementary charge, $\epsilon$ the charge carrier energy, $v(\epsilon)^2 = {2\epsilon}/{m^*}$ the group velocity squared, $m^*$ the carrier effective mass, $\tau(\epsilon)$ the relaxation time, $f(\epsilon) = [e^{(\epsilon-\mu)/k_B T}+1]^{-1}$ the Fermi-Dirac distribution, $\mu$ the chemical potential, and
$
g(\epsilon) = \sqrt{2}\pi^{-2} \hbar^{-3}({m^*})^{3/2} \epsilon^{1/2} 
$
the electronic density of states. 
We consider donor doping by phosphorous (activation energy 45 meV) at a concentration of $3 \times 10^{19}$ cm$^{-3}$. 
The resulting carrier density and Fermi level are determined self-consistently {\it via} a graphical iteration method \cite{kittel2005introduction}.
The relaxation time $\tau(\epsilon)$ remains the only unknown to be determined.

To determine $\tau(\epsilon)$, we consider intrinsic and extrinsic scattering processes, the latter arising here directly from the damaged regions created by ion beam irradiation. 
Matthiessen's law gives the overall scattering rate as 
\begin{equation} \label{Eq_RlxT}
\tau_0^{-1}(\epsilon)=\tau_{i}^{-1}(\epsilon)+\tau_{D}^{-1}(\epsilon)  \hspace{0.5em},
\end{equation} 
where $\tau_i(\epsilon)$ denotes intrinsic and $\tau_D(\epsilon)$ extrinsic scattering times. 
This amounts to assuming that the defective regions act as isolated scattering centers. 
For $\tau_i(\epsilon)$, we assume that in the irradiated samples the intrinsic scattering mechanisms remain unchanged from pristine silicon  \cite{lundstrom2009fundamentals}, a commonly used assumption when studying nanotructured thermoelectric metamaterials \cite{lee2008nanoporous}. 
We incorporate descriptions of intrinsic electron scattering according to the deformation potential of acoustic phonons and optical phonons; all the material parameters and models are summarized in Table \ref{tabScat}. 
Scattering rates for both acoustic and optical phonons share the power-law form  $\tau_i(\epsilon) = \tau_{i0} (\epsilon/k_B T)^r$, where the parameters $\tau_{i0}$ and $r$ can be fitted to experimental measurements, and have previously been well characterized for silicon \cite{jacoboni2010theory}. 

\begin{table*}[h]
\caption{The scattering mechanisms and corresponding power-law models, $\tau_i(\epsilon) = \tau_{i0} x^{r_i}, x= \epsilon/k_B T$, considered in this work. The dominant scattering mechanisms around and above room temperature are deformation potential scattering with acoustic and optical phonons. The parameters are obtained by fitting experimental measurements:\cite{jacoboni2010theory,wolfe1988physical} $D_A=9.0$ eV, $C_l=(3C_{11}+2C_{12}+4C_{44})/5= 1.895 \times 10^7$ Pa, $\theta=\hbar \omega_{LO}/k_B=731.1$ K. Note that the unified power-law with identical exponents $r_i$ largely simplify the analysis in this work.
}
\label{tabScat}
\begin{tabular}{l c c c }
\hline
\hline
Scattering mechanism ($i$) & $\tau_{i0}$  &  $r_i$  & Refs.  \\\hline 
\parbox{4cm}{Acoustic phonon deformation potential} &  
\parbox{7cm}{ \begin{equation}  \nonumber
\frac{2.40 \times 10^{-19} C_l }{D_A^2 T^{3/2}} \left( \frac{m}{m^*}\right)^{3/2}
\end{equation}} &  
\parbox{2cm}{$-1/2$} & 
[\onlinecite{jacoboni2010theory}, \onlinecite{wolfe1988physical}]  \\
\parbox{4cm}{Optical phonon deformation potential} & 
\parbox{6cm}{ \begin{equation}  \nonumber
\frac{4.83 \times 10^{-19} C_l [\exp (\theta/T)-1]}{D_A^2 T^{1/2} \theta} \left( \frac{m}{m^*}\right)^{3/2}
\end{equation}}  & 
$-1/2$ & 
[\onlinecite{jacoboni2010theory}, \onlinecite{wolfe1988physical}]  \\
Cylindrical defective area & 
\parbox{7cm}{ \begin{equation}  \nonumber
\frac{\pi}{ 4 \sqrt{2}}\frac{ L^2 }{D} \sqrt{\frac{m^*}{k_B T}}
\end{equation}} & $-1/2$ & \parbox{2cm}{Eqn. \ref{eqTauD} in this work}
\\
\hline \hline
\end{tabular}
\end{table*}

On the other hand, in order to determine $\tau_{D}(\epsilon)$, we invoked the partial wave approach. 
Partial wave analysis is a general method to calculate scattering cross-sections applicable when the scattering potential is azimuthally symmetric, \cite{schiff1968quantum} which is an approximate but reasonable description of the ion beam damaged regions. 
This approach has been applied recently to estimate the scattering time for electrons interacting with spherical quantum dots embedded in a host matrix. \cite{zebarjadi2009effect}
In the following section, we adapt the method to cylindrical, rather than spherical, defective regions of interest here. 
This theoretical scattering model will also be applicable to other recently proposed planar-patterned nanomaterials\cite{graczykowski2017thermal} and two-dimensional nanoporous/holey metamaterials \cite{xu2014holey,lin2015holey}.

\subsection{Relaxation time $\tau_D(\epsilon)$ due to cylindrical defects}
\begin{figure*}[!hbtp] 
\centering
\includegraphics[width=0.9\textwidth]{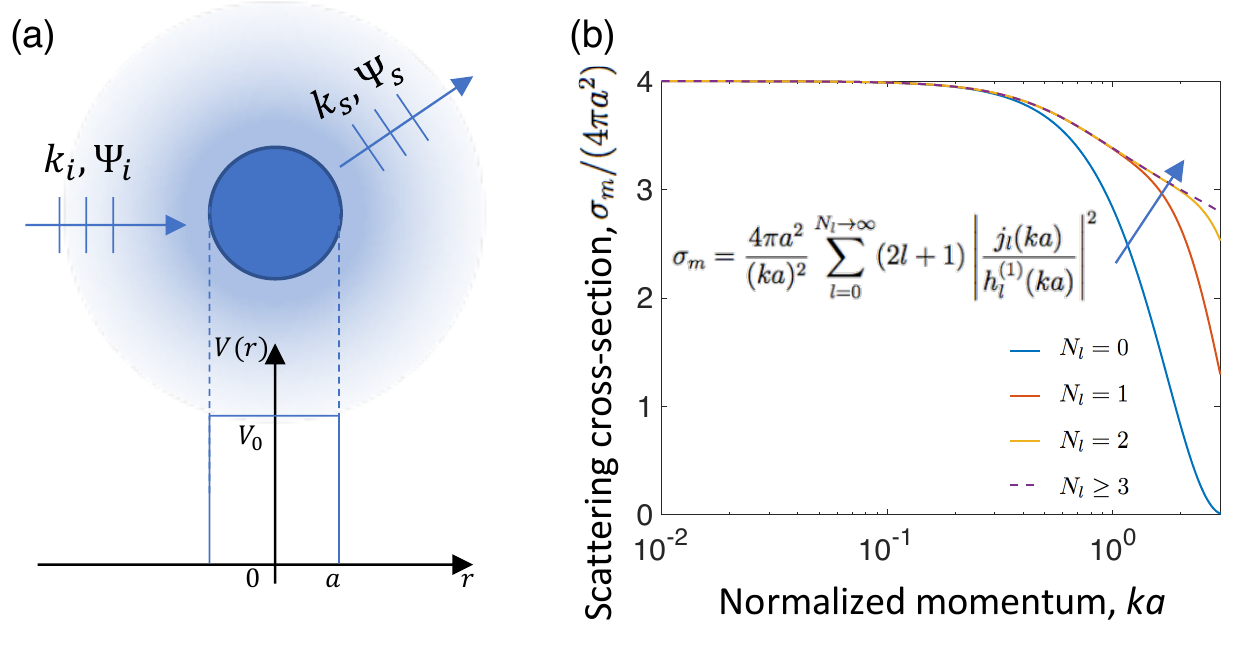}
\caption{\label{figTauD}  (a) Schematic of the scattering model for calculating the scattering cross-section and scattering time with cylindrical defective area. (b) The convergence of the exact solution to infinite scattering barrier with the number of partial waves $N_l$ considered. For $ka<0.5$ the $S$-wave ($N_l=0, l=0$) is sufficient for accurate descriptions of the scattering cross-section.
}
\end{figure*}

In the following we derive the scattering rate for electrons $\tau_{D}(\epsilon)$ due to the presence of a  cylindrical barrier potential, as shown in Fig. \ref{figTauD}(a),
\begin{equation} \label{eq_Pot}
V(r)=\begin{cases}
	V_0 ,& ~r \leq a\\
	~0 ,& ~r>a
 \end{cases}
\end{equation}
where $V_0>0$ is the barrier height. Assuming the scattering is elastic, kinetic theory gives
\begin{equation} \label{eqScatR}
\tau_{D}(\epsilon)^{-1}=N_D \left\langle v \right\rangle D_m   \hspace{0.5em}, 
\end{equation}
where $N_D$ is the density of defected regions, $\left\langle v \right\rangle$ the average carrier velocity, $D_m=\sqrt[]{4\sigma_m/\pi}$ is the scattering diameter, and $\sigma_m$ denotes the momentum scattering cross-section defined by
\begin{equation} \label{eqSigDef}
\sigma_m=\int\sigma(\theta)(1-\cos\theta) d \Omega=2 \pi \int_0^\pi \sigma(\theta) (1-\cos \theta) \sin \theta d \theta   \hspace{0.5em},
\end{equation}
where $\sigma(\theta)=\frac{d\sigma}{d\Omega}$ is the differential scattering cross-section that measures the probability of incident particles passing through an infinitesimal area $d\sigma$ and then being scattered into  solid angle $d\Omega$. Here the differential cross-section is independent of azimuthal angle due to the potential symmetry. 

A detailed derivation of the scattering cross section using partial wave analysis is provided in the Appendix \ref{appndSigm}. 
In the limit of low energy elastic scattering process, the cross-section is
\begin{equation} \label{eqSigM0}
\sigma_m \approx 4 \pi a^2 \left(1-\frac{\tanh(k_0 a)}{k_0 a}\right)^2 \approx 4 \pi a^2  \hspace{0.5em}, 
\end{equation}
which is an approximate solution obtained by retaining only $S$-wave ($l=0$) component of the complete solution
\begin{equation}
\sigma_m=\frac{4 \pi a^2}{(ka)^2} \sum_{l=0}^{N_l \rightarrow \infty} (2l+1) \left | \frac{j_l(ka)}{h_l^{(1)}(ka)} \right |^2 ,
\end{equation}
where $j_l$ and $h_l^{(1)}$ are the spherical Bessel and first-kind Hankel functions, $k^2=2m \epsilon/\hbar^2$, and $k_0^2=2mV_0/\hbar^2$. 
An {\it a posteriori} justification of the assumed $S$-wave scattering, with higher-order terms neglected, is presented in Fig. \ref{figTauD}(b). In the limit of an insulating, impermeable defective region ($V_0 \rightarrow \infty$), the boundary condition becomes $\psi(a,\theta) = 0$. 
As seen from Fig. \ref{figTauD}(b), the calculated  cross-section converges quickly with the number of angular terms ($N_l$) included. For instance when $ka=0.5$ with only $l=0$, an error of 1.91\% is introduced. 
Therefore, retaining the $l=0$ term alone well represents low-energy scattering ($ka\ll 1$). 

Before substituting Eqn. \ref{eqSigM0} into Eqn. \ref{eqScatR} to obtain the scattering rate, the average velocity of incident carriers must be found. 
Within the parabolic band description adopted here, the carrier speed is related to the energy as $v=\sqrt[]{2 \epsilon/m^*}$. 
Due to the uniform distribution of angles $\vartheta \in [-\pi/2, \pi/2]$ between the velocity vector and the longitudinal cylinder axis, the average incident speed is
\begin{equation} \label{eqVavg}
\braket{v}=\int_{-\pi/2}^{\pi/2} v \cos \vartheta \Theta (\vartheta) d \vartheta=\frac{2v}{\pi} ,
\end{equation}
with the distribution density $\Theta (\vartheta)=1/\pi$. 

Combining Eqns. \ref{eqScatR}, \ref{eqSigM0}, and \ref{eqVavg}, and letting $N_D=1/L^2$ be the number density of the defective areas, the momentum relaxation time can be written as
\begin{equation} \label{eqTauD}
\tau_{D}(\epsilon)^{-1}=N_D \braket{v} D_m = \frac{4 \sqrt{2}}{ \pi}\frac{ D }{L^2} \sqrt{\frac{k_B T}{m^*}} x^{1/2}   \hspace{0.5em},
\end{equation}
where $x=\epsilon/k_B T$.
Ultimately, the external scattering rate due to the cylindrical defective areas exhibits the power-law form  $\tau_D(\epsilon) = \tau_{D0} (\epsilon/k_B T)^r$ with exponent $r=1/2$, which turns out to be the same scaling as all intrinsic models (See Table \ref{tabScat}).
The unified power-law scattering conveniently simplifies our analysis, allowing a unified calculation of electrical properties. 
Substituting $\tau(\epsilon) = (\tau_{i0}+\tau_{D0}) (\epsilon/k_B T)^r$, $r=1/2$ into Eqns. (\ref{sigmaBoltz}) and (\ref{SBoltz}), 
\begin{widetext}
\begin{eqnarray}
\label{sigma}
\sigma &=& \frac{2q^2 \tau_0 (3/2+r) (k_B T)^{3/2+r} \Gamma(3/2+r)}{3 \sqrt[]{2} \pi^{3/2} \Gamma(3/2)}(m^*)^{1/2} e^\eta  \hspace{0.5em}, \\
\label{S}
S &=& -\frac{k_B}{q}(\eta-r-\frac{5}{2})  \hspace{0.5em}, 
\end{eqnarray}
\end{widetext} 
where $\Gamma$ denotes the gamma function and $\eta=\mu/(k_B T)$ the reduced chemical potential.

\section{Results \& discussion}
\subsection{Thermoelectric properties of defective silicon metamaterials}

\begin{figure*}[!hbtp] 
\centering
\includegraphics[width=0.8\textwidth]{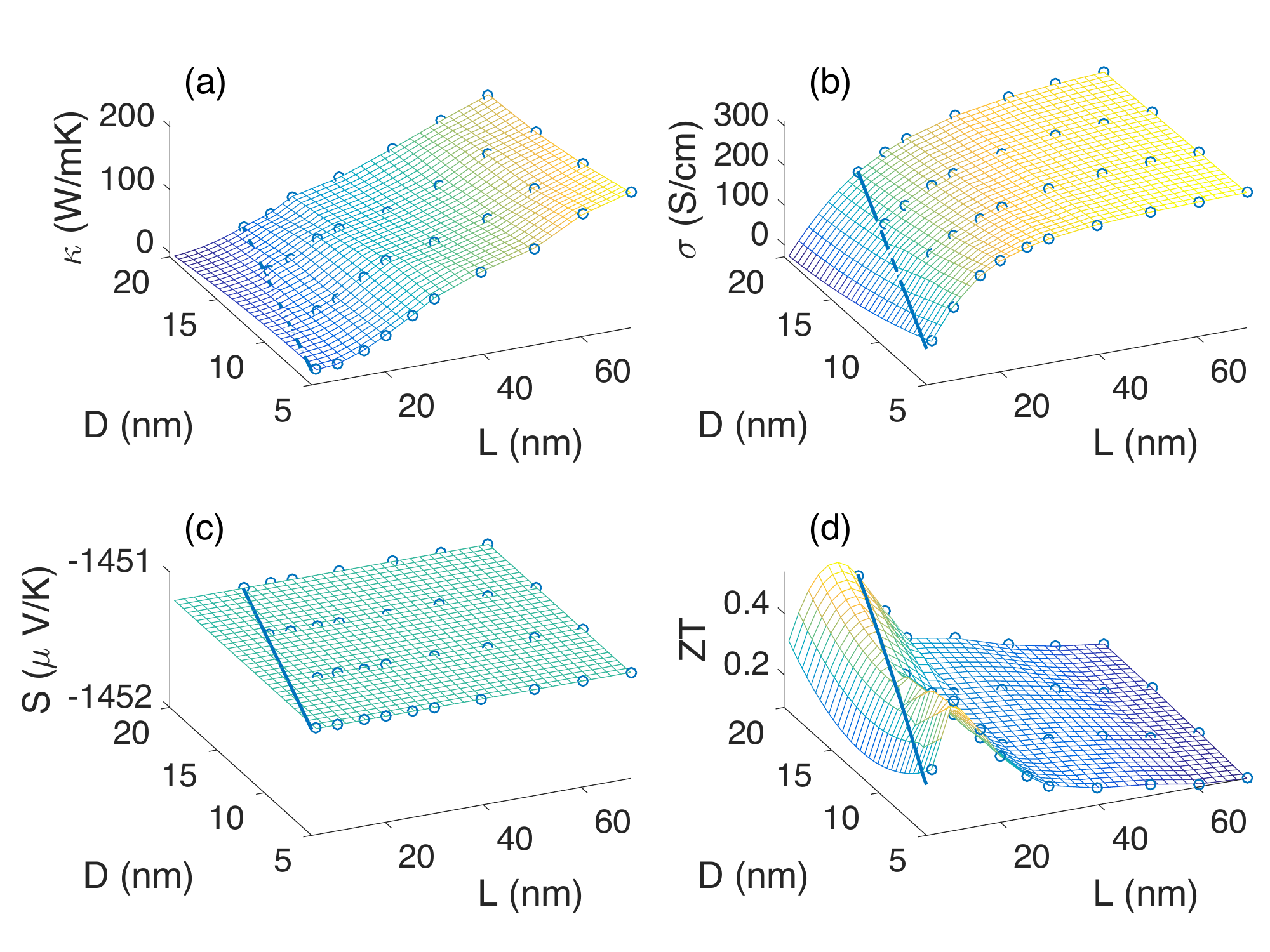}
\caption{\label{figZT}  (a) Thermal conductivity $\kappa$, (b) electrical conductivity $\sigma$, (c) Seebeck coefficient $S$, and (d) figure of merit $ZT$ as function of $D$ and $L$ for $n$-type silicon doped at a concentration of $3\times 10^{19}$ cm$^{-3}$ at room temperature. 
Both $\sigma$ and $\kappa$ are more sensitive to $L$ than $D$.
Electron conductivity $\sigma$ grows quickly and saturates sooner than thermal conductivity $\kappa$ with $L$, which allows the patterned metamaterial to be more crystalline for electrons than for phonons. 
For $T=300$K at the given dopant concentration, $ZT$ can be enhanced to around $ZT \approx 0.5$ for optimal $L \approx 11$~nm, $D \approx 5$~nm. The thick blue lines correspond to $L=D$; the region to the left of the lines are geometrically unphysical.
}
\end{figure*}

Using the equilibrium molecular dynamics simulations, we predict $\kappa$ as a function of the geometric parameters  $D$ and $L$, as summarized in Fig. \ref{figZT}(a). 
The thermal conductivity of the irradiated metamaterials is suppressed appreciably compared to pristine silicon. For instance, with $L=11$ nm and $D=5$ nm, $\kappa$ is reduced by a factor of 19 from $270$ W/mK for crystalline silicon. 
In our forthcoming work, combining lattice dynamics and molecular dynamics, this reduction in $\kappa$ is found to arise largely from hybridization, interactions, and avoided crossings between bulk-like vibrational modes and modes confined to the defective regions\cite{TZEE2017}. 
As $L$ increases, $\kappa$ is expected to approach the numerical value of $270$ W/mK for bulk silicon.
The lattice conductivity in Fig. \ref{figZT}(a) shows a large sensitivity to the interdefect distance $L$, and is less sensitive to the defect diameter $D$ (discussed further below).

The electrical properties $\sigma$ and $S$ are plotted similarly as functions of $L$ and $D$ in Fig. \ref{figZT}(b,c) from the closed form expressions in Eqs. (\ref{sigma}) and (\ref{S}). 
From Fig. \ref{figZT}(b), we notice that $\sigma$ is also more sensitive to $L$ than $D$, similar to $\kappa$ in Fig. \ref{figZT}(a). 
Furthermore, $\sigma$ is observed to increase sharply with $L$ when $L<20$ nm, but starts to saturate to the bulk value for larger $L$. 
The contrast between the slow, smooth drop for $\kappa$ in Fig. \ref{figZT}(a) across the full range of $L$, and the sharper collapse for $\sigma$ in Fig. \ref{figZT}(b) for $L < 20$~nm results in a window where the PGEC concept of Fig. \ref{figSchem}(b) can be realized. 
To better understand these trends, we provide a scaling analysis of $\kappa,\sigma$ with $D,L$ in the following section. 
Meanwhile, from Fig. \ref{figZT}(c) the Seebeck coefficient $S$ is not affected by the variations of $L$ and $D$ in the classical model  used here. 
This can be understood from Eq. (\ref{S}), which shows that $S$ depends only on the reduced Fermi level and the scattering mechanisms. 
Since $r=-1/2$ for both electron-phonon and electron-defect scattering, for a given dopant concentration and temperature, the reduced Fermi level is fixed and $S$ is independent of the absolute scattering time and thus the defect density.

When combined together, the thermal and electrical properties in Figs. \ref{figZT}(a-c) lead to a figure of merit $ZT$ as shown in Fig. \ref{figZT}(d). 
As $L$ decreases, $ZT$ can be enhanced 18 fold compared to bulk silicon, reaching as high as $ZT \approx 0.5$ for $L \approx 11$~nm, $D \approx 5$~nm. 
This value may even underestimate the actual attainable $ZT$ by nearly a factor of two, since $\kappa$ is overestimated by the same amount using the Tersoff potential. 
In the silicon metamaterial both electrical and thermal conductivities are reduced by the patterned defective regions, but $ZT$ is set by the ratio of electrical to thermal properties, rather than their individual absolute values. 
For small $L$, $\sigma$ grows faster than $\kappa$ and the material is more ``crystalline'' for electrons than for phonons. 
Therefore, as surmised, the regularly patterned defects can achieve a PGEC  with $\Lambda_e < L < \Lambda_p$. 


\subsection{Sensitivity of thermoelectric properties to $L$ and $D$}
\label{senDL}
In this section, we present a scaling analysis to understand both the greater sensitivity of $\kappa$ and $\sigma$ to $L$ than $D$, and the more rapid recovery of $\sigma$ than $\kappa$ as $L$ increases. 
Both $\kappa$ and $\sigma$ can be written as a function of $D$ and $L$,
\begin{equation}
\zeta(D,L)=b(D,L) \Lambda(D,L)  \hspace{0.5em}, 
\end{equation}
where $\zeta=$ $\kappa$ or $\sigma$, $b(D,L)$ accounts for the changes in band structure for both phonons and electrons, and $\Lambda(D,L)$ is the mean free path. 
In the following, we assume the band function $b(D,L)$ is constant, insensitive to $D$ and $L$, which is accurate when $D \ll L$, or $D$ and $L$ vary in a narrow range, as considered in this work.

Therefore, the sensitivity can be defined as
\begin{equation} \label{C2}
\frac{\partial \zeta}{\partial(D,L)}=\frac{\partial \zeta}{\partial \Lambda} \frac{\partial \Lambda}{\partial (D,L)}  \hspace{0.5em},
\end{equation}
where $\partial(D,L)$ denotes partial derivative with respect to $D$ or $L$.
Similar to Eq. \ref{Eq_RlxT}, Matthiessen's law for mean free path can be written as
\begin{equation} \label{C3}
\Lambda(D,L)=\frac{\Lambda_i \Lambda_D (D,L)}{\Lambda_i + \Lambda_D (D,L)}  \hspace{0.5em}. 
\end{equation}
Note that $\Lambda_i$ represents the intrinsic mean free path in pristine silicon and is assumed  insensitive to $(D,L)$. 
Substituting Eq. \ref{C3} into Eq. \ref{C2},
\begin{equation} \label{C4}
\frac{\partial \zeta}{\partial(D,L)} =
\left( \frac{\Lambda_i}{\Lambda_i + \Lambda_D} \right)^2  \frac{\partial \Lambda_D}{\partial(D,L)}  \hspace{0.5em}. 
\end{equation}
Applying Eq. \ref{eqTauD} of the main text for $\tau_D$,
\begin{eqnarray}
\frac{\partial \zeta}{\partial D} &=&
-b(D,L) \left( \frac{\Lambda_i}{\Lambda_i + L^2/2D} \right)^2 \frac{L^2}{2D^2}    \hspace{0.5em}, \\
\frac{\partial \zeta}{\partial L} &=&
b(D,L) \left( \frac{\Lambda_i}{\Lambda_i + L^2/2D} \right)^2 \frac{L}{D}     \hspace{0.5em}.
\end{eqnarray}

\begin{figure*}[!hbtp]
\centering
\includegraphics[width=0.9\textwidth]{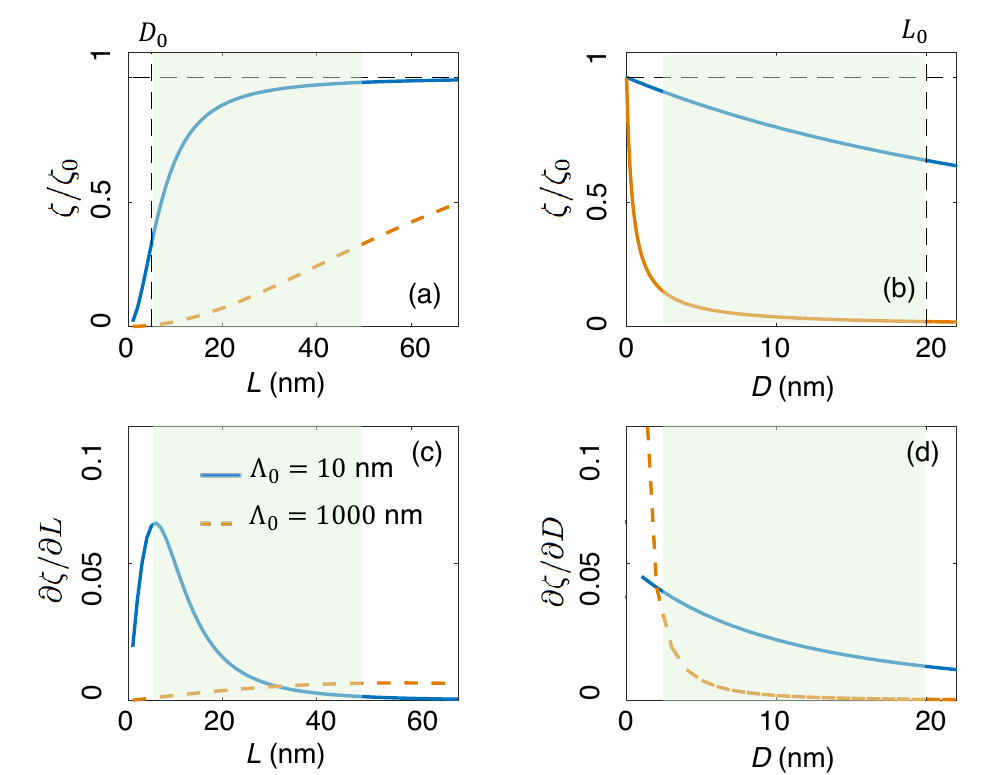}
\caption{
\label{figSensi} 
Scaling of $\zeta=\kappa$ or $\sigma$ as functions of (a) $L$ and (b) $D$ based on kinetic arguments for different intrinsic mean free paths $\Lambda_0$. While $\Lambda_0=10$ nm represents the scaling of electron conductivity, $\Lambda_0=1000$ nm approximates the trend of phonon conductivity. For direct comparison the values of $L$ and $D$ are given with dimensions. The shaded areas signify the parameter ranges considered in this work. (c, d) Sensitivity to $L$ and $D$. In the range of $L$, $\kappa$ keeps increasing and $\sigma$ saturates earlier. Meanwhile, for the considered range of $D$, the conductivities are similarly sensitive to $D$ and $L$. 
These results are consistent with the trends in Fig. \ref{figZT}(a,b).  
}
\end{figure*}

These scaling forms and corresponding sensitivity are shown in Fig. \ref{figSensi}. 
Two sets of results are shown, for intrinsic mean free paths $\Lambda_0=10$ nm and $\Lambda_0=1000$ nm. 
The former represents $\Lambda_e$, while the latter $\Lambda_p$, in silicon. 
In the relevant ranges of $D$ and $L$, we observe similar sensitivity of $\sigma$ and $\kappa$ to $L$ and $D$. 
This scaling analysis also recovers the early saturation in $\sigma$ for $L > 20$ nm compared to $\kappa$. 
These trends are consistent with those in Fig. \ref{figZT}(a,b).

\section{conclusion}
We showed that regularly patterned nanoscale defects formed by ion beam irradiation in silicon can be used to realize a phonon glass electron crystal, of interest for thermoelectric applications. 
When the distance between the patterned defects lies within the length window of electron and phonon mean free paths, the thermal conductivity can be reduced without substantial detriment to the electrical properties. 
Using the Green-Kubo relations and equilibrium molecular dynamics, we predict a 19 fold reduction in $\kappa$. 
Meanwhile, with Boltzmann theory the electrical power factor is shown to retain more than 80\% of its value in crystalline silicon. 
To apply Boltzmann theory we use partial wave analysis to derive a scattering model for electrons in a cylindrical potential. 
Combining these predictions,  we obtain a $ZT \approx 0.5$ or greater at room temperature. 
In consideration of economic and manufacturing aspects, silicon has been chosen as a representative material. However the physical trends observed may apply to other materials as well, particularly those with longer phonon mean free paths. 

\section*{Acknowledgment} 
We gratefully acknowledge Jun Ma and Emil Annevelink from Illinois for helpful discussions. 
We acknowledge financial support from the National Science Foundation under Grant No. EAGER-1550895. 
Computational resources were provided by both (i) the Blue Waters sustained petascale computing facilities, and (ii) the Illinois Campus Computing Cluster.

\newpage  

\newpage 


\newpage 

\newpage 

\newpage

\appendix
\section{Scattering cross-section from partial wave analysis}
\label{appndSigm}
For the azimuthally symmetric potential in Eqn. \ref{eq_Pot}, an incident plane wave $\psi_i(z)=A \exp{(i\mathbf{k}\cdot\mathbf{r})}$ is expected to be scattered into a spherical wave (see, for example, [\onlinecite{schiff1968quantum}]),
\begin{equation} \label{eqScatW}
\psi_S(r)=A f(k,\theta) \frac{\exp (ikr)}{r}  \hspace{0.5em}, 
\end{equation}
where $f(k,\theta)$ is the scattering amplitude, and a composite wave field
\begin{equation}
\psi(r) = \psi_i(r)+\psi_S(r)   \hspace{0.5em} 
\end{equation}
should be sought as the solution to the Schr\"{o}dinger equation
\begin{equation} \label{eqTISE}
\left[ \nabla^2 + k^2 - \frac{2 m}{\hbar^2} V(r) \right] \psi(r) =0 \hspace{1em}, 
\end{equation}
where $k^2=2m \epsilon /\hbar^2$. The time-independent form is employed since the scattering is assumed to be elastic and thus energy remains unchanged during scattering.

The probability of the incident particle with speed $v$ passing through an infinitesimal area $d\sigma$ in time $dt$ is $dP=|A|^2~v~dt~d\sigma$, which is equal to the probability of scattering into the corresponding solid angle $d\Omega$,  
$dP=|A|^2~|f(k,\theta)|^2~v~dt~r^2~d\Omega$. Thus, by definition the differential scattering cross-section is 
\begin{equation}
\sigma_m(\theta)=|f(k,\theta)|^2  \hspace{0.5em}. 
\end{equation}
Therefore, to determine the scattering rate $\tau_D^{-1}$ in Eqn. \ref{eqScatR}, we need only to calculate the scattering amplitude $f(k,\theta)$ in Eqn. \ref{eqScatW}. For this, two possible methods are partial wave analysis and the Born approximation. However, the latter assumes a small scattering potential so that the scattering field is only slightly changed from the incident wave field.  
Since the ion beam patterned regions are expected to introduce substantial scattering, it is necessary to consider large scattering barriers for which the Born approximation becomes singular. 
Therefore, we derive the scattering cross-section and momentum relaxation using partial wave expansion which remains valid. 


The partial wave method decomposes the incident and scattered wavefunctions into partial spherical waves, and then imposes boundary conditions to determine the partial wave magnitudes or phase shifts for each (see, for example, Ref. [\onlinecite{schiff1968quantum}]). Based on partial wave analysis for the azimuthally symmetric potential, the differential cross-section is formulated as
\begin{equation}
\sigma(\theta)=\frac{1}{k^2} \left | \sum_{l=0}^{\infty} (2l+1) e^{i \delta_l} \sin \delta_l P_l(\cos \theta) \right |^2  \hspace{0.5em},
\end{equation}
where $\delta_l$ is the phase shift between incident and scattered waves, and $P_l$ is the $l$th Legendre polynomial. The scattering process can be completely determined if the phase shifts $\delta_l$ are known for all partial waves. However, this method is particularly useful when dealing with low-energy scattering ($ka \ll 1$), where only the first term ($l=0$, the so-called $S$-wave) dominates. We consider in this work $S$-wave scattering, which is also consistent with the assumption of isotropic scattering as required by Boltzmann theory. \cite{wolfe1988physical,jacoboni2010theory} In other words, based on the definition in Eqn. \ref{eqSigDef},
\begin{equation} \label{eqSig0}
\sigma_m=\frac{4 \pi}{k^2} \sin^2 \delta_0  \hspace{0.5em}.
\end{equation}

The solution of the Sch\"{o}dinger equation (Eqn. \ref{eqTISE}) thus formulated is separable, and the radial components of the equation are 
\begin{equation}
\begin{cases}
	\frac{d u^2}{d r^2} + (k^2-k_0^2) u = 0, & r\leq a \\
	\frac{d u^2}{d r^2} + k^2 u = 0, & r > a
 \end{cases} \hspace{1em},
\end{equation}
where $k^2=2m \epsilon/\hbar^2$ and $k_0^2=2mV_0/\hbar^2$. The solutions are 
\begin{equation}
u(r)=\begin{cases}
	A \sinh(k_1 r), & r\leq a \\
	B \sin(kr+ \delta_0), & r > a
 \end{cases} \hspace{1em},
\end{equation}
where $k_1^2=k_0^2-k^2$. Imposing the continuity of wave functions and their derivatives at $r=a$ gives  
\begin{subequations}
\begin{align}
A \sinh(k_1 a) &= B \sin(kr+ \delta_0) \hspace{0.5em},  \\
A k_1 \cosh(k_1 a) &= B k \cos(kr+ \delta_0)
 \hspace{0.5em}. 
\end{align}
\end{subequations}
Dividing the two equations above, we obtain
\begin{equation} \label{eqDelt0}
\tan \delta_0=\frac{k \tanh (k_1 a) -k_1 \tan (k a)}{k_1  + k \tan (k a) \tanh (k_1 a)} \hspace{0.5em}.
\end{equation}
Using Eqns. \ref{eqSig0} and \ref{eqDelt0}, the scattering cross-section can be determined as 
\begin{equation}
\sigma_m=\frac{4 \pi a^2}{(ka)^2} \sum_{l=0}^{N_l \rightarrow \infty} (2l+1) \left | \frac{j_l(ka)}{h_l^{(1)}(ka)} \right |^2 ,
\end{equation}
where $j_l$ and $h_l^{(1)}$ are the spherical Bessel and first-kind Hankel functions. 
In the limits of low carrier energy $ka \ll 1$ and high barrier $k_1 a \gg 1$, we have $k_1 a \approx k_0 a$ and the above equation can be  simplified to
\begin{equation}
\tan \delta_0 \approx \delta_0  \approx k\left( \frac{\tanh (k_0 a) -k_0 a}{k_0} \right) \approx  -ka  \hspace{0.5em}, 
\end{equation}
and
\begin{equation}
\sigma_m \approx 4 \pi a^2 \left(1-\frac{\tanh(k_0 a)}{k_0 a}\right)^2 \approx 4 \pi a^2  \hspace{0.5em}, 
\end{equation}
which is Eqn. \ref{eqSigM0} in the main text.

\newpage  
\bibliography{main}

\begin{thebibliography}{41}%
\makeatletter
\providecommand \@ifxundefined [1]{%
 \@ifx{#1\undefined}
}%
\providecommand \@ifnum [1]{%
 \ifnum #1\expandafter \@firstoftwo
 \else \expandafter \@secondoftwo
 \fi
}%
\providecommand \@ifx [1]{%
 \ifx #1\expandafter \@firstoftwo
 \else \expandafter \@secondoftwo
 \fi
}%
\providecommand \natexlab [1]{#1}%
\providecommand \enquote  [1]{``#1''}%
\providecommand \bibnamefont  [1]{#1}%
\providecommand \bibfnamefont [1]{#1}%
\providecommand \citenamefont [1]{#1}%
\providecommand \href@noop [0]{\@secondoftwo}%
\providecommand \href [0]{\begingroup \@sanitize@url \@href}%
\providecommand \@href[1]{\@@startlink{#1}\@@href}%
\providecommand \@@href[1]{\endgroup#1\@@endlink}%
\providecommand \@sanitize@url [0]{\catcode `\\12\catcode `\$12\catcode
  `\&12\catcode `\#12\catcode `\^12\catcode `\_12\catcode `\%12\relax}%
\providecommand \@@startlink[1]{}%
\providecommand \@@endlink[0]{}%
\providecommand \url  [0]{\begingroup\@sanitize@url \@url }%
\providecommand \@url [1]{\endgroup\@href {#1}{\urlprefix }}%
\providecommand \urlprefix  [0]{URL }%
\providecommand \Eprint [0]{\href }%
\providecommand \doibase [0]{http://dx.doi.org/}%
\providecommand \selectlanguage [0]{\@gobble}%
\providecommand \bibinfo  [0]{\@secondoftwo}%
\providecommand \bibfield  [0]{\@secondoftwo}%
\providecommand \translation [1]{[#1]}%
\providecommand \BibitemOpen [0]{}%
\providecommand \bibitemStop [0]{}%
\providecommand \bibitemNoStop [0]{.\EOS\space}%
\providecommand \EOS [0]{\spacefactor3000\relax}%
\providecommand \BibitemShut  [1]{\csname bibitem#1\endcsname}%
\let\auto@bib@innerbib\@empty
\bibitem [{\citenamefont {Majumdar}(2004)}]{majumdar2004thermoelectricity}%
  \BibitemOpen
  \bibfield  {author} {\bibinfo {author} {\bibfnamefont {A.}~\bibnamefont
  {Majumdar}},\ }\href@noop {} {\bibfield  {journal} {\bibinfo  {journal}
  {Science}\ }\textbf {\bibinfo {volume} {303}},\ \bibinfo {pages} {777}
  (\bibinfo {year} {2004})}\BibitemShut {NoStop}%
\bibitem [{\citenamefont {Tritt}\ and\ \citenamefont
  {Subramanian}(2006)}]{tritt2006thermoelectric}%
  \BibitemOpen
  \bibfield  {author} {\bibinfo {author} {\bibfnamefont {T.~M.}\ \bibnamefont
  {Tritt}}\ and\ \bibinfo {author} {\bibfnamefont {M.}~\bibnamefont
  {Subramanian}},\ }\href@noop {} {\bibfield  {journal} {\bibinfo  {journal}
  {MRS Bulletin}\ }\textbf {\bibinfo {volume} {31}},\ \bibinfo {pages} {188}
  (\bibinfo {year} {2006})}\BibitemShut {NoStop}%
\bibitem [{\citenamefont {Sootsman}\ \emph {et~al.}(2009)\citenamefont
  {Sootsman}, \citenamefont {Chung},\ and\ \citenamefont
  {Kanatzidis}}]{sootsman2009new}%
  \BibitemOpen
  \bibfield  {author} {\bibinfo {author} {\bibfnamefont {J.~R.}\ \bibnamefont
  {Sootsman}}, \bibinfo {author} {\bibfnamefont {D.~Y.}\ \bibnamefont {Chung}},
  \ and\ \bibinfo {author} {\bibfnamefont {M.~G.}\ \bibnamefont {Kanatzidis}},\
  }\href@noop {} {\bibfield  {journal} {\bibinfo  {journal} {Angewandte Chemie
  International Edition}\ }\textbf {\bibinfo {volume} {48}},\ \bibinfo {pages}
  {8616} (\bibinfo {year} {2009})}\BibitemShut {NoStop}%
\bibitem [{\citenamefont {Minnich}\ \emph {et~al.}(2009)\citenamefont
  {Minnich}, \citenamefont {Dresselhaus}, \citenamefont {Ren},\ and\
  \citenamefont {Chen}}]{minnich2009bulk}%
  \BibitemOpen
  \bibfield  {author} {\bibinfo {author} {\bibfnamefont {A.}~\bibnamefont
  {Minnich}}, \bibinfo {author} {\bibfnamefont {M.}~\bibnamefont
  {Dresselhaus}}, \bibinfo {author} {\bibfnamefont {Z.}~\bibnamefont {Ren}}, \
  and\ \bibinfo {author} {\bibfnamefont {G.}~\bibnamefont {Chen}},\ }\href@noop
  {} {\bibfield  {journal} {\bibinfo  {journal} {Energy \& Environmental
  Science}\ }\textbf {\bibinfo {volume} {2}},\ \bibinfo {pages} {466} (\bibinfo
  {year} {2009})}\BibitemShut {NoStop}%
\bibitem [{\citenamefont {Heremans}\ \emph {et~al.}(2013)\citenamefont
  {Heremans}, \citenamefont {Dresselhaus}, \citenamefont {Bell},\ and\
  \citenamefont {Morelli}}]{heremans2013thermoelectrics}%
  \BibitemOpen
  \bibfield  {author} {\bibinfo {author} {\bibfnamefont {J.~P.}\ \bibnamefont
  {Heremans}}, \bibinfo {author} {\bibfnamefont {M.~S.}\ \bibnamefont
  {Dresselhaus}}, \bibinfo {author} {\bibfnamefont {L.~E.}\ \bibnamefont
  {Bell}}, \ and\ \bibinfo {author} {\bibfnamefont {D.~T.}\ \bibnamefont
  {Morelli}},\ }\href@noop {} {\bibfield  {journal} {\bibinfo  {journal}
  {Nature Nanotechnology}\ }\textbf {\bibinfo {volume} {8}},\ \bibinfo {pages}
  {471} (\bibinfo {year} {2013})}\BibitemShut {NoStop}%
\bibitem [{\citenamefont {Nolas}\ \emph {et~al.}(2013)\citenamefont {Nolas},
  \citenamefont {Sharp},\ and\ \citenamefont
  {Goldsmid}}]{nolas2013thermoelectrics}%
  \BibitemOpen
  \bibfield  {author} {\bibinfo {author} {\bibfnamefont {G.~S.}\ \bibnamefont
  {Nolas}}, \bibinfo {author} {\bibfnamefont {J.}~\bibnamefont {Sharp}}, \ and\
  \bibinfo {author} {\bibfnamefont {J.}~\bibnamefont {Goldsmid}},\ }\href@noop
  {} {\emph {\bibinfo {title} {Thermoelectrics: basic principles and new
  materials developments}}},\ Vol.~\bibinfo {volume} {45}\ (\bibinfo
  {publisher} {Springer Science \& Business Media},\ \bibinfo {year}
  {2013})\BibitemShut {NoStop}%
\bibitem [{\citenamefont {Balandin}(2005)}]{balandin2005nanophononics}%
  \BibitemOpen
  \bibfield  {author} {\bibinfo {author} {\bibfnamefont {A.~A.}\ \bibnamefont
  {Balandin}},\ }\href@noop {} {\bibfield  {journal} {\bibinfo  {journal}
  {Journal of Nanoscience and Nanotechnology}\ }\textbf {\bibinfo {volume}
  {5}},\ \bibinfo {pages} {1015} (\bibinfo {year} {2005})}\BibitemShut
  {NoStop}%
\bibitem [{\citenamefont {Hicks}\ and\ \citenamefont
  {Dresselhaus}(1993)}]{hicks1993effect}%
  \BibitemOpen
  \bibfield  {author} {\bibinfo {author} {\bibfnamefont {L.}~\bibnamefont
  {Hicks}}\ and\ \bibinfo {author} {\bibfnamefont {M.}~\bibnamefont
  {Dresselhaus}},\ }\href@noop {} {\bibfield  {journal} {\bibinfo  {journal}
  {Physical Review B}\ }\textbf {\bibinfo {volume} {47}},\ \bibinfo {pages}
  {12727} (\bibinfo {year} {1993})}\BibitemShut {NoStop}%
\bibitem [{\citenamefont {Dresselhaus}\ \emph {et~al.}(2007)\citenamefont
  {Dresselhaus}, \citenamefont {Chen}, \citenamefont {Tang}, \citenamefont
  {Yang}, \citenamefont {Lee}, \citenamefont {Wang}, \citenamefont {Ren},
  \citenamefont {Fleurial},\ and\ \citenamefont {Gogna}}]{dresselhaus2007new}%
  \BibitemOpen
  \bibfield  {author} {\bibinfo {author} {\bibfnamefont {M.~S.}\ \bibnamefont
  {Dresselhaus}}, \bibinfo {author} {\bibfnamefont {G.}~\bibnamefont {Chen}},
  \bibinfo {author} {\bibfnamefont {M.~Y.}\ \bibnamefont {Tang}}, \bibinfo
  {author} {\bibfnamefont {R.}~\bibnamefont {Yang}}, \bibinfo {author}
  {\bibfnamefont {H.}~\bibnamefont {Lee}}, \bibinfo {author} {\bibfnamefont
  {D.}~\bibnamefont {Wang}}, \bibinfo {author} {\bibfnamefont {Z.}~\bibnamefont
  {Ren}}, \bibinfo {author} {\bibfnamefont {J.-P.}\ \bibnamefont {Fleurial}}, \
  and\ \bibinfo {author} {\bibfnamefont {P.}~\bibnamefont {Gogna}},\
  }\href@noop {} {\bibfield  {journal} {\bibinfo  {journal} {Advanced
  Materials}\ }\textbf {\bibinfo {volume} {19}},\ \bibinfo {pages} {1043}
  (\bibinfo {year} {2007})}\BibitemShut {NoStop}%
\bibitem [{\citenamefont {Slack}(1995)}]{slackPGEC1995}%
  \BibitemOpen
  \bibfield  {author} {\bibinfo {author} {\bibfnamefont {G.~A.}\ \bibnamefont
  {Slack}},\ }in\ \href@noop {} {\emph {\bibinfo {booktitle} {CRC handbook of
  thermoelectrics}}},\ \bibinfo {editor} {edited by\ \bibinfo {editor}
  {\bibfnamefont {D.~M.}\ \bibnamefont {Rowe}}}\ (\bibinfo  {publisher} {CRC
  press},\ \bibinfo {address} {Boca Raton, FL},\ \bibinfo {year} {1995})\
  Chap.~\bibinfo {chapter} {34}, pp.\ \bibinfo {pages} {407--440}\BibitemShut
  {NoStop}%
\bibitem [{\citenamefont {Liu}\ \emph {et~al.}(2013)\citenamefont {Liu},
  \citenamefont {Yuan}, \citenamefont {Lu}, \citenamefont {Shi}, \citenamefont
  {Xu}, \citenamefont {He}, \citenamefont {Tang}, \citenamefont {Bai},
  \citenamefont {Zhang}, \citenamefont {Chen} \emph
  {et~al.}}]{liu2013ultrahigh}%
  \BibitemOpen
  \bibfield  {author} {\bibinfo {author} {\bibfnamefont {H.}~\bibnamefont
  {Liu}}, \bibinfo {author} {\bibfnamefont {X.}~\bibnamefont {Yuan}}, \bibinfo
  {author} {\bibfnamefont {P.}~\bibnamefont {Lu}}, \bibinfo {author}
  {\bibfnamefont {X.}~\bibnamefont {Shi}}, \bibinfo {author} {\bibfnamefont
  {F.}~\bibnamefont {Xu}}, \bibinfo {author} {\bibfnamefont {Y.}~\bibnamefont
  {He}}, \bibinfo {author} {\bibfnamefont {Y.}~\bibnamefont {Tang}}, \bibinfo
  {author} {\bibfnamefont {S.}~\bibnamefont {Bai}}, \bibinfo {author}
  {\bibfnamefont {W.}~\bibnamefont {Zhang}}, \bibinfo {author} {\bibfnamefont
  {L.}~\bibnamefont {Chen}},  \emph {et~al.},\ }\href@noop {} {\bibfield
  {journal} {\bibinfo  {journal} {Advanced Materials}\ }\textbf {\bibinfo
  {volume} {25}},\ \bibinfo {pages} {6607} (\bibinfo {year}
  {2013})}\BibitemShut {NoStop}%
\bibitem [{\citenamefont {Venkatasubramanian}\ \emph
  {et~al.}(2001)\citenamefont {Venkatasubramanian}, \citenamefont {Siivola},
  \citenamefont {Colpitts},\ and\ \citenamefont
  {O'quinn}}]{venkatasubramanian2001thin}%
  \BibitemOpen
  \bibfield  {author} {\bibinfo {author} {\bibfnamefont {R.}~\bibnamefont
  {Venkatasubramanian}}, \bibinfo {author} {\bibfnamefont {E.}~\bibnamefont
  {Siivola}}, \bibinfo {author} {\bibfnamefont {T.}~\bibnamefont {Colpitts}}, \
  and\ \bibinfo {author} {\bibfnamefont {B.}~\bibnamefont {O'quinn}},\
  }\href@noop {} {\bibfield  {journal} {\bibinfo  {journal} {Nature}\ }\textbf
  {\bibinfo {volume} {413}},\ \bibinfo {pages} {597} (\bibinfo {year}
  {2001})}\BibitemShut {NoStop}%
\bibitem [{\citenamefont {Harman}\ \emph {et~al.}(2002)\citenamefont {Harman},
  \citenamefont {Taylor}, \citenamefont {Walsh},\ and\ \citenamefont
  {LaForge}}]{harman2002quantum}%
  \BibitemOpen
  \bibfield  {author} {\bibinfo {author} {\bibfnamefont {T.}~\bibnamefont
  {Harman}}, \bibinfo {author} {\bibfnamefont {P.}~\bibnamefont {Taylor}},
  \bibinfo {author} {\bibfnamefont {M.}~\bibnamefont {Walsh}}, \ and\ \bibinfo
  {author} {\bibfnamefont {B.}~\bibnamefont {LaForge}},\ }\href@noop {}
  {\bibfield  {journal} {\bibinfo  {journal} {Science}\ }\textbf {\bibinfo
  {volume} {297}},\ \bibinfo {pages} {2229} (\bibinfo {year}
  {2002})}\BibitemShut {NoStop}%
\bibitem [{\citenamefont {Snyder}\ and\ \citenamefont
  {Toberer}(2008)}]{Snyder2008}%
  \BibitemOpen
  \bibfield  {author} {\bibinfo {author} {\bibfnamefont {G.~J.}\ \bibnamefont
  {Snyder}}\ and\ \bibinfo {author} {\bibfnamefont {E.~S.}\ \bibnamefont
  {Toberer}},\ }\href {\doibase 10.1038/nmat2090} {\bibfield  {journal}
  {\bibinfo  {journal} {Nature materials}\ }\textbf {\bibinfo {volume} {7}},\
  \bibinfo {pages} {105} (\bibinfo {year} {2008})}\BibitemShut {NoStop}%
\bibitem [{\citenamefont {Nolas}\ \emph {et~al.}(1999)\citenamefont {Nolas},
  \citenamefont {Morelli},\ and\ \citenamefont
  {Tritt}}]{nolas1999skutterudites}%
  \BibitemOpen
  \bibfield  {author} {\bibinfo {author} {\bibfnamefont {G.}~\bibnamefont
  {Nolas}}, \bibinfo {author} {\bibfnamefont {D.}~\bibnamefont {Morelli}}, \
  and\ \bibinfo {author} {\bibfnamefont {T.~M.}\ \bibnamefont {Tritt}},\
  }\href@noop {} {\bibfield  {journal} {\bibinfo  {journal} {Annual Review of
  Materials Science}\ }\textbf {\bibinfo {volume} {29}},\ \bibinfo {pages} {89}
  (\bibinfo {year} {1999})}\BibitemShut {NoStop}%
\bibitem [{\citenamefont {Zhu}\ \emph {et~al.}(2015)\citenamefont {Zhu},
  \citenamefont {Fu}, \citenamefont {Xie}, \citenamefont {Liu},\ and\
  \citenamefont {Zhao}}]{zhu2015high}%
  \BibitemOpen
  \bibfield  {author} {\bibinfo {author} {\bibfnamefont {T.}~\bibnamefont
  {Zhu}}, \bibinfo {author} {\bibfnamefont {C.}~\bibnamefont {Fu}}, \bibinfo
  {author} {\bibfnamefont {H.}~\bibnamefont {Xie}}, \bibinfo {author}
  {\bibfnamefont {Y.}~\bibnamefont {Liu}}, \ and\ \bibinfo {author}
  {\bibfnamefont {X.}~\bibnamefont {Zhao}},\ }\href@noop {} {\bibfield
  {journal} {\bibinfo  {journal} {Advanced Energy Materials}\ }\textbf
  {\bibinfo {volume} {5}} (\bibinfo {year} {2015})}\BibitemShut {NoStop}%
\bibitem [{\citenamefont {LeBlanc}\ \emph {et~al.}(2014)\citenamefont
  {LeBlanc}, \citenamefont {Yee}, \citenamefont {Scullin}, \citenamefont
  {Dames},\ and\ \citenamefont {Goodson}}]{leblanc2014material}%
  \BibitemOpen
  \bibfield  {author} {\bibinfo {author} {\bibfnamefont {S.}~\bibnamefont
  {LeBlanc}}, \bibinfo {author} {\bibfnamefont {S.~K.}\ \bibnamefont {Yee}},
  \bibinfo {author} {\bibfnamefont {M.~L.}\ \bibnamefont {Scullin}}, \bibinfo
  {author} {\bibfnamefont {C.}~\bibnamefont {Dames}}, \ and\ \bibinfo {author}
  {\bibfnamefont {K.~E.}\ \bibnamefont {Goodson}},\ }\href@noop {} {\bibfield
  {journal} {\bibinfo  {journal} {Renewable and Sustainable Energy Reviews}\
  }\textbf {\bibinfo {volume} {32}},\ \bibinfo {pages} {313} (\bibinfo {year}
  {2014})}\BibitemShut {NoStop}%
\bibitem [{\citenamefont {Hochbaum}\ \emph {et~al.}(2008)\citenamefont
  {Hochbaum}, \citenamefont {Chen}, \citenamefont {Delgado}, \citenamefont
  {Liang}, \citenamefont {Garnett}, \citenamefont {Najarian}, \citenamefont
  {Majumdar},\ and\ \citenamefont {Yang}}]{hochbaum2008enhanced}%
  \BibitemOpen
  \bibfield  {author} {\bibinfo {author} {\bibfnamefont {A.~I.}\ \bibnamefont
  {Hochbaum}}, \bibinfo {author} {\bibfnamefont {R.}~\bibnamefont {Chen}},
  \bibinfo {author} {\bibfnamefont {R.~D.}\ \bibnamefont {Delgado}}, \bibinfo
  {author} {\bibfnamefont {W.}~\bibnamefont {Liang}}, \bibinfo {author}
  {\bibfnamefont {E.~C.}\ \bibnamefont {Garnett}}, \bibinfo {author}
  {\bibfnamefont {M.}~\bibnamefont {Najarian}}, \bibinfo {author}
  {\bibfnamefont {A.}~\bibnamefont {Majumdar}}, \ and\ \bibinfo {author}
  {\bibfnamefont {P.}~\bibnamefont {Yang}},\ }\href@noop {} {\bibfield
  {journal} {\bibinfo  {journal} {Nature}\ }\textbf {\bibinfo {volume} {451}},\
  \bibinfo {pages} {163} (\bibinfo {year} {2008})}\BibitemShut {NoStop}%
\bibitem [{\citenamefont {Bux}\ \emph {et~al.}(2009)\citenamefont {Bux},
  \citenamefont {Blair}, \citenamefont {Gogna}, \citenamefont {Lee},
  \citenamefont {Chen}, \citenamefont {Dresselhaus}, \citenamefont {Kaner},\
  and\ \citenamefont {Fleurial}}]{bux2009nanostructured}%
  \BibitemOpen
  \bibfield  {author} {\bibinfo {author} {\bibfnamefont {S.~K.}\ \bibnamefont
  {Bux}}, \bibinfo {author} {\bibfnamefont {R.~G.}\ \bibnamefont {Blair}},
  \bibinfo {author} {\bibfnamefont {P.~K.}\ \bibnamefont {Gogna}}, \bibinfo
  {author} {\bibfnamefont {H.}~\bibnamefont {Lee}}, \bibinfo {author}
  {\bibfnamefont {G.}~\bibnamefont {Chen}}, \bibinfo {author} {\bibfnamefont
  {M.~S.}\ \bibnamefont {Dresselhaus}}, \bibinfo {author} {\bibfnamefont
  {R.~B.}\ \bibnamefont {Kaner}}, \ and\ \bibinfo {author} {\bibfnamefont
  {J.-P.}\ \bibnamefont {Fleurial}},\ }\href@noop {} {\bibfield  {journal}
  {\bibinfo  {journal} {Advanced Functional Materials}\ }\textbf {\bibinfo
  {volume} {19}},\ \bibinfo {pages} {2445} (\bibinfo {year}
  {2009})}\BibitemShut {NoStop}%
\bibitem [{\citenamefont {Lee}\ \emph {et~al.}(2008)\citenamefont {Lee},
  \citenamefont {Galli},\ and\ \citenamefont {Grossman}}]{lee2008nanoporous}%
  \BibitemOpen
  \bibfield  {author} {\bibinfo {author} {\bibfnamefont {J.-H.}\ \bibnamefont
  {Lee}}, \bibinfo {author} {\bibfnamefont {G.~A.}\ \bibnamefont {Galli}}, \
  and\ \bibinfo {author} {\bibfnamefont {J.~C.}\ \bibnamefont {Grossman}},\
  }\href@noop {} {\bibfield  {journal} {\bibinfo  {journal} {Nano Letters}\
  }\textbf {\bibinfo {volume} {8}},\ \bibinfo {pages} {3750} (\bibinfo {year}
  {2008})}\BibitemShut {NoStop}%
\bibitem [{\citenamefont {Yee}\ \emph {et~al.}(2013)\citenamefont {Yee},
  \citenamefont {LeBlanc}, \citenamefont {Goodson},\ and\ \citenamefont
  {Dames}}]{yee2013per}%
  \BibitemOpen
  \bibfield  {author} {\bibinfo {author} {\bibfnamefont {S.~K.}\ \bibnamefont
  {Yee}}, \bibinfo {author} {\bibfnamefont {S.}~\bibnamefont {LeBlanc}},
  \bibinfo {author} {\bibfnamefont {K.~E.}\ \bibnamefont {Goodson}}, \ and\
  \bibinfo {author} {\bibfnamefont {C.}~\bibnamefont {Dames}},\ }\href@noop {}
  {\bibfield  {journal} {\bibinfo  {journal} {Energy \& Environmental Science}\
  }\textbf {\bibinfo {volume} {6}},\ \bibinfo {pages} {2561} (\bibinfo {year}
  {2013})}\BibitemShut {NoStop}%
\bibitem [{\citenamefont {Swaminathan-Gopalan}\ \emph
  {et~al.}(2017)\citenamefont {Swaminathan-Gopalan}, \citenamefont {Zhu},
  \citenamefont {Ertekin},\ and\ \citenamefont {Stephani}}]{defects2016}%
  \BibitemOpen
  \bibfield  {author} {\bibinfo {author} {\bibfnamefont {K.}~\bibnamefont
  {Swaminathan-Gopalan}}, \bibinfo {author} {\bibfnamefont {T.}~\bibnamefont
  {Zhu}}, \bibinfo {author} {\bibfnamefont {E.}~\bibnamefont {Ertekin}}, \ and\
  \bibinfo {author} {\bibfnamefont {K.}~\bibnamefont {Stephani}},\ }\href@noop
  {} {\bibfield  {journal} {\bibinfo  {journal} {Physics Review B}\ }\textbf
  {\bibinfo {volume} {95}},\ \bibinfo {pages} {184109} (\bibinfo {year}
  {2017})}\BibitemShut {NoStop}%
\bibitem [{\citenamefont {Boukai}\ \emph {et~al.}(2008)\citenamefont {Boukai},
  \citenamefont {Bunimovich}, \citenamefont {Tahir-Kheli}, \citenamefont {Yu},
  \citenamefont {Goddard~Iii},\ and\ \citenamefont
  {Heath}}]{boukai2008silicon}%
  \BibitemOpen
  \bibfield  {author} {\bibinfo {author} {\bibfnamefont {A.~I.}\ \bibnamefont
  {Boukai}}, \bibinfo {author} {\bibfnamefont {Y.}~\bibnamefont {Bunimovich}},
  \bibinfo {author} {\bibfnamefont {J.}~\bibnamefont {Tahir-Kheli}}, \bibinfo
  {author} {\bibfnamefont {J.-K.}\ \bibnamefont {Yu}}, \bibinfo {author}
  {\bibfnamefont {W.~A.}\ \bibnamefont {Goddard~Iii}}, \ and\ \bibinfo {author}
  {\bibfnamefont {J.~R.}\ \bibnamefont {Heath}},\ }\href@noop {} {\bibfield
  {journal} {\bibinfo  {journal} {Nature}\ }\textbf {\bibinfo {volume} {451}},\
  \bibinfo {pages} {168} (\bibinfo {year} {2008})}\BibitemShut {NoStop}%
\bibitem [{\citenamefont {Tang}\ \emph {et~al.}(2010)\citenamefont {Tang},
  \citenamefont {Wang}, \citenamefont {Lee}, \citenamefont {Fardy},
  \citenamefont {Huo}, \citenamefont {Russell},\ and\ \citenamefont
  {Yang}}]{tang2010holey}%
  \BibitemOpen
  \bibfield  {author} {\bibinfo {author} {\bibfnamefont {J.}~\bibnamefont
  {Tang}}, \bibinfo {author} {\bibfnamefont {H.-T.}\ \bibnamefont {Wang}},
  \bibinfo {author} {\bibfnamefont {D.~H.}\ \bibnamefont {Lee}}, \bibinfo
  {author} {\bibfnamefont {M.}~\bibnamefont {Fardy}}, \bibinfo {author}
  {\bibfnamefont {Z.}~\bibnamefont {Huo}}, \bibinfo {author} {\bibfnamefont
  {T.~P.}\ \bibnamefont {Russell}}, \ and\ \bibinfo {author} {\bibfnamefont
  {P.}~\bibnamefont {Yang}},\ }\href@noop {} {\bibfield  {journal} {\bibinfo
  {journal} {Nano Letters}\ }\textbf {\bibinfo {volume} {10}},\ \bibinfo
  {pages} {4279} (\bibinfo {year} {2010})}\BibitemShut {NoStop}%
\bibitem [{\citenamefont {Anderson}\ \emph {et~al.}(2008)\citenamefont
  {Anderson}, \citenamefont {Lorenz},\ and\ \citenamefont
  {Travesset}}]{Anderson2008}%
  \BibitemOpen
  \bibfield  {author} {\bibinfo {author} {\bibfnamefont {J.~A.}\ \bibnamefont
  {Anderson}}, \bibinfo {author} {\bibfnamefont {C.~D.}\ \bibnamefont
  {Lorenz}}, \ and\ \bibinfo {author} {\bibfnamefont {A.}~\bibnamefont
  {Travesset}},\ }\href {\doibase 10.1016/j.jcp.2008.01.047} {\bibfield
  {journal} {\bibinfo  {journal} {Journal of Computational Physics}\ }\textbf
  {\bibinfo {volume} {227}},\ \bibinfo {pages} {5342} (\bibinfo {year}
  {2008})}\BibitemShut {NoStop}%
\bibitem [{\citenamefont {Tersoff}(1988)}]{Tersoff1988}%
  \BibitemOpen
  \bibfield  {author} {\bibinfo {author} {\bibfnamefont {J.}~\bibnamefont
  {Tersoff}},\ }\href {\doibase 10.1103/PhysRevB.38.9902} {\bibfield  {journal}
  {\bibinfo  {journal} {Physical Review B}\ }\textbf {\bibinfo {volume} {38}},\
  \bibinfo {pages} {9902} (\bibinfo {year} {1988})}\BibitemShut {NoStop}%
\bibitem [{\citenamefont {Ziegler}\ \emph {et~al.}(1985)\citenamefont
  {Ziegler}, \citenamefont {Biersack},\ and\ \citenamefont
  {Littmark}}]{ziegler1985stopping}%
  \BibitemOpen
  \bibfield  {author} {\bibinfo {author} {\bibfnamefont {J.~F.}\ \bibnamefont
  {Ziegler}}, \bibinfo {author} {\bibfnamefont {J.}~\bibnamefont {Biersack}}, \
  and\ \bibinfo {author} {\bibfnamefont {U.}~\bibnamefont {Littmark}},\
  }\href@noop {} {\emph {\bibinfo {title} {The stopping and range of ions in
  matter, Vol. 1}}}\ (\bibinfo  {publisher} {Pergamon, New York},\ \bibinfo
  {year} {1985})\BibitemShut {NoStop}%
\bibitem [{\citenamefont {Frenkel}\ and\ \citenamefont
  {Smit}(2001)}]{frenkel2001MD}%
  \BibitemOpen
  \bibfield  {author} {\bibinfo {author} {\bibfnamefont {D.}~\bibnamefont
  {Frenkel}}\ and\ \bibinfo {author} {\bibfnamefont {B.}~\bibnamefont {Smit}},\
  }\href@noop {} {\emph {\bibinfo {title} {Understanding molecular simulation:
  from algorithms to applications}}},\ Vol.~\bibinfo {volume} {1}\ (\bibinfo
  {publisher} {Academic press},\ \bibinfo {year} {2001})\BibitemShut {NoStop}%
\bibitem [{\citenamefont {Fulkerson}\ \emph {et~al.}(1968)\citenamefont
  {Fulkerson}, \citenamefont {Moore}, \citenamefont {Williams}, \citenamefont
  {Graves},\ and\ \citenamefont {McElroy}}]{Fulkerson1968}%
  \BibitemOpen
  \bibfield  {author} {\bibinfo {author} {\bibfnamefont {W.}~\bibnamefont
  {Fulkerson}}, \bibinfo {author} {\bibfnamefont {J.}~\bibnamefont {Moore}},
  \bibinfo {author} {\bibfnamefont {R.}~\bibnamefont {Williams}}, \bibinfo
  {author} {\bibfnamefont {R.}~\bibnamefont {Graves}}, \ and\ \bibinfo {author}
  {\bibfnamefont {D.}~\bibnamefont {McElroy}},\ }\href@noop {} {\bibfield
  {journal} {\bibinfo  {journal} {Physical Review}\ }\textbf {\bibinfo {volume}
  {167}},\ \bibinfo {pages} {765} (\bibinfo {year} {1968})}\BibitemShut
  {NoStop}%
\bibitem [{\citenamefont {Lee}\ \emph {et~al.}(2007)\citenamefont {Lee},
  \citenamefont {Grossman}, \citenamefont {Reed},\ and\ \citenamefont
  {Galli}}]{lee2007lattice}%
  \BibitemOpen
  \bibfield  {author} {\bibinfo {author} {\bibfnamefont {J.}~\bibnamefont
  {Lee}}, \bibinfo {author} {\bibfnamefont {J.}~\bibnamefont {Grossman}},
  \bibinfo {author} {\bibfnamefont {J.}~\bibnamefont {Reed}}, \ and\ \bibinfo
  {author} {\bibfnamefont {G.}~\bibnamefont {Galli}},\ }\href@noop {}
  {\bibfield  {journal} {\bibinfo  {journal} {Applied Physics Letters}\
  }\textbf {\bibinfo {volume} {91}},\ \bibinfo {pages} {223110} (\bibinfo
  {year} {2007})}\BibitemShut {NoStop}%
\bibitem [{\citenamefont {Lundstrom}(2009)}]{lundstrom2009fundamentals}%
  \BibitemOpen
  \bibfield  {author} {\bibinfo {author} {\bibfnamefont {M.}~\bibnamefont
  {Lundstrom}},\ }\href@noop {} {\emph {\bibinfo {title} {Fundamentals of
  carrier transport}}}\ (\bibinfo  {publisher} {Cambridge University Press},\
  \bibinfo {year} {2009})\BibitemShut {NoStop}%
\bibitem [{\citenamefont {Ma}\ and\ \citenamefont
  {Sinha}(2012)}]{ma2012thermoelectric}%
  \BibitemOpen
  \bibfield  {author} {\bibinfo {author} {\bibfnamefont {J.}~\bibnamefont
  {Ma}}\ and\ \bibinfo {author} {\bibfnamefont {S.}~\bibnamefont {Sinha}},\
  }\href@noop {} {\bibfield  {journal} {\bibinfo  {journal} {Journal of Applied
  Physics}\ }\textbf {\bibinfo {volume} {112}},\ \bibinfo {pages} {073719}
  (\bibinfo {year} {2012})}\BibitemShut {NoStop}%
\bibitem [{\citenamefont {Kittel}(2005)}]{kittel2005introduction}%
  \BibitemOpen
  \bibfield  {author} {\bibinfo {author} {\bibfnamefont {C.}~\bibnamefont
  {Kittel}},\ }\href@noop {} {\emph {\bibinfo {title} {Introduction to solid
  state physics}}}\ (\bibinfo  {publisher} {John Wiley \& Sons},\ \bibinfo
  {year} {2005})\BibitemShut {NoStop}%
\bibitem [{\citenamefont {Jacoboni}(2010)}]{jacoboni2010theory}%
  \BibitemOpen
  \bibfield  {author} {\bibinfo {author} {\bibfnamefont {C.}~\bibnamefont
  {Jacoboni}},\ }\href@noop {} {\emph {\bibinfo {title} {Theory of Electron
  Transport in Semiconductors: A Pathway from Elementary Physics to
  Nonequilibrium Green Functions}}},\ Vol.\ \bibinfo {volume} {165}\ (\bibinfo
  {publisher} {Springer Science \& Business Media},\ \bibinfo {year}
  {2010})\BibitemShut {NoStop}%
\bibitem [{\citenamefont {Wolfe}\ \emph {et~al.}(1988)\citenamefont {Wolfe},
  \citenamefont {Holonyak~Jr},\ and\ \citenamefont
  {Stillman}}]{wolfe1988physical}%
  \BibitemOpen
  \bibfield  {author} {\bibinfo {author} {\bibfnamefont {C.~M.}\ \bibnamefont
  {Wolfe}}, \bibinfo {author} {\bibfnamefont {N.}~\bibnamefont {Holonyak~Jr}},
  \ and\ \bibinfo {author} {\bibfnamefont {G.~E.}\ \bibnamefont {Stillman}},\
  }\href@noop {} {\emph {\bibinfo {title} {Physical properties of
  semiconductors}}}\ (\bibinfo  {publisher} {Prentice-Hall, Inc.},\ \bibinfo
  {year} {1988})\BibitemShut {NoStop}%
\bibitem [{\citenamefont {Schiff}(1968)}]{schiff1968quantum}%
  \BibitemOpen
  \bibfield  {author} {\bibinfo {author} {\bibfnamefont {L.~I.}\ \bibnamefont
  {Schiff}},\ }\href@noop {} {\emph {\bibinfo {title} {Quantum mechanics}}}\
  (\bibinfo  {publisher} {McGraw-Hill},\ \bibinfo {year} {1968})\BibitemShut
  {NoStop}%
\bibitem [{\citenamefont {Zebarjadi}\ \emph {et~al.}(2009)\citenamefont
  {Zebarjadi}, \citenamefont {Esfarjani}, \citenamefont {Shakouri},
  \citenamefont {Bahk}, \citenamefont {Bian}, \citenamefont {Zeng},
  \citenamefont {Bowers}, \citenamefont {Lu}, \citenamefont {Zide},\ and\
  \citenamefont {Gossard}}]{zebarjadi2009effect}%
  \BibitemOpen
  \bibfield  {author} {\bibinfo {author} {\bibfnamefont {M.}~\bibnamefont
  {Zebarjadi}}, \bibinfo {author} {\bibfnamefont {K.}~\bibnamefont
  {Esfarjani}}, \bibinfo {author} {\bibfnamefont {A.}~\bibnamefont {Shakouri}},
  \bibinfo {author} {\bibfnamefont {J.-H.}\ \bibnamefont {Bahk}}, \bibinfo
  {author} {\bibfnamefont {Z.}~\bibnamefont {Bian}}, \bibinfo {author}
  {\bibfnamefont {G.}~\bibnamefont {Zeng}}, \bibinfo {author} {\bibfnamefont
  {J.}~\bibnamefont {Bowers}}, \bibinfo {author} {\bibfnamefont
  {H.}~\bibnamefont {Lu}}, \bibinfo {author} {\bibfnamefont {J.}~\bibnamefont
  {Zide}}, \ and\ \bibinfo {author} {\bibfnamefont {A.}~\bibnamefont
  {Gossard}},\ }\href@noop {} {\bibfield  {journal} {\bibinfo  {journal}
  {Applied Physics Letters}\ }\textbf {\bibinfo {volume} {94}},\ \bibinfo
  {pages} {202105} (\bibinfo {year} {2009})}\BibitemShut {NoStop}%
\bibitem [{\citenamefont {Graczykowski}\ \emph {et~al.}(2017)\citenamefont
  {Graczykowski}, \citenamefont {El~Sachat}, \citenamefont {Reparaz},
  \citenamefont {Sledzinska}, \citenamefont {Wagner}, \citenamefont
  {Chavez-Angel}, \citenamefont {Wu}, \citenamefont {Volz}, \citenamefont {Wu},
  \citenamefont {Alzina} \emph {et~al.}}]{graczykowski2017thermal}%
  \BibitemOpen
  \bibfield  {author} {\bibinfo {author} {\bibfnamefont {B.}~\bibnamefont
  {Graczykowski}}, \bibinfo {author} {\bibfnamefont {A.}~\bibnamefont
  {El~Sachat}}, \bibinfo {author} {\bibfnamefont {J.}~\bibnamefont {Reparaz}},
  \bibinfo {author} {\bibfnamefont {M.}~\bibnamefont {Sledzinska}}, \bibinfo
  {author} {\bibfnamefont {M.}~\bibnamefont {Wagner}}, \bibinfo {author}
  {\bibfnamefont {E.}~\bibnamefont {Chavez-Angel}}, \bibinfo {author}
  {\bibfnamefont {Y.}~\bibnamefont {Wu}}, \bibinfo {author} {\bibfnamefont
  {S.}~\bibnamefont {Volz}}, \bibinfo {author} {\bibfnamefont {Y.}~\bibnamefont
  {Wu}}, \bibinfo {author} {\bibfnamefont {F.}~\bibnamefont {Alzina}},  \emph
  {et~al.},\ }\href@noop {} {\bibfield  {journal} {\bibinfo  {journal} {Nature
  Communications}\ }\textbf {\bibinfo {volume} {8}} (\bibinfo {year}
  {2017})}\BibitemShut {NoStop}%
\bibitem [{\citenamefont {Xu}\ \emph {et~al.}(2014)\citenamefont {Xu},
  \citenamefont {Lin}, \citenamefont {Zhong}, \citenamefont {Huang},
  \citenamefont {Weiss}, \citenamefont {Huang},\ and\ \citenamefont
  {Duan}}]{xu2014holey}%
  \BibitemOpen
  \bibfield  {author} {\bibinfo {author} {\bibfnamefont {Y.}~\bibnamefont
  {Xu}}, \bibinfo {author} {\bibfnamefont {Z.}~\bibnamefont {Lin}}, \bibinfo
  {author} {\bibfnamefont {X.}~\bibnamefont {Zhong}}, \bibinfo {author}
  {\bibfnamefont {X.}~\bibnamefont {Huang}}, \bibinfo {author} {\bibfnamefont
  {N.~O.}\ \bibnamefont {Weiss}}, \bibinfo {author} {\bibfnamefont
  {Y.}~\bibnamefont {Huang}}, \ and\ \bibinfo {author} {\bibfnamefont
  {X.}~\bibnamefont {Duan}},\ }\href@noop {} {\bibfield  {journal} {\bibinfo
  {journal} {Nature communications}\ }\textbf {\bibinfo {volume} {5}},\
  \bibinfo {pages} {4554} (\bibinfo {year} {2014})}\BibitemShut {NoStop}%
\bibitem [{\citenamefont {Lin}\ \emph {et~al.}(2015)\citenamefont {Lin},
  \citenamefont {Han}, \citenamefont {Campbell}, \citenamefont {Kim},
  \citenamefont {Zhao}, \citenamefont {Luo}, \citenamefont {Dai}, \citenamefont
  {Hu},\ and\ \citenamefont {Connell}}]{lin2015holey}%
  \BibitemOpen
  \bibfield  {author} {\bibinfo {author} {\bibfnamefont {Y.}~\bibnamefont
  {Lin}}, \bibinfo {author} {\bibfnamefont {X.}~\bibnamefont {Han}}, \bibinfo
  {author} {\bibfnamefont {C.~J.}\ \bibnamefont {Campbell}}, \bibinfo {author}
  {\bibfnamefont {J.-W.}\ \bibnamefont {Kim}}, \bibinfo {author} {\bibfnamefont
  {B.}~\bibnamefont {Zhao}}, \bibinfo {author} {\bibfnamefont {W.}~\bibnamefont
  {Luo}}, \bibinfo {author} {\bibfnamefont {J.}~\bibnamefont {Dai}}, \bibinfo
  {author} {\bibfnamefont {L.}~\bibnamefont {Hu}}, \ and\ \bibinfo {author}
  {\bibfnamefont {J.~W.}\ \bibnamefont {Connell}},\ }\href@noop {} {\bibfield
  {journal} {\bibinfo  {journal} {Advanced Functional Materials}\ }\textbf
  {\bibinfo {volume} {25}},\ \bibinfo {pages} {2920} (\bibinfo {year}
  {2015})}\BibitemShut {NoStop}%
\bibitem [{\citenamefont {Zhu}\ \emph {et~al.}()\citenamefont {Zhu},
  \citenamefont {Swaminathan-Gopalan}, \citenamefont {Stephani},\ and\
  \citenamefont {Ertekin}}]{TZEE2017}%
  \BibitemOpen
  \bibfield  {author} {\bibinfo {author} {\bibfnamefont {T.}~\bibnamefont
  {Zhu}}, \bibinfo {author} {\bibfnamefont {K.}~\bibnamefont
  {Swaminathan-Gopalan}}, \bibinfo {author} {\bibfnamefont {K.}~\bibnamefont
  {Stephani}}, \ and\ \bibinfo {author} {\bibfnamefont {E.}~\bibnamefont
  {Ertekin}},\ }\href@noop {} {\bibinfo  {journal} {submitted}\ }\BibitemShut
  {NoStop}%
\end{thebibliography}%

\end{document}